# Novel Carbon allotropes with mixed hybridizations: *ene*-C$_{10}$, and *ene-yne*-C$_{14}$. Crystal chemistry and first principles investigations.

Samir F. Matar

Lebanese German University (LGU), Sahel Alma, Jounieh, Lebanon

https://orcid.org/0000-0001-5419-358X

Email: s.matar@lgu.edu.lb & abouliess@gmail.com

*Abstract*

*Based on C$_8$ "carbon 4C" with **cfc** topology, two hybrid carbon allotropes generated by inserting C(sp$^2$) and C(sp$^1$) carbon atoms into C$_8$ diamond-like lattice were identified and labeled ene-C$_{10}$ containing C(sp$^2$) and ene-yne-C$_{14}$ containing C(sp$^2$ and sp$^1$). The introduced double C=C and triple C≡C chemical descriptions were illustrated from the projected charge densities. The crystal density and the cohesive energy were found to decrease due to the enhanced openness of the structures from inserted sp$^2$/sp$^1$ carbons, with a resulting decrease of the hardness along the series C$_8$, C$_{10}$, C$_{12}$, and C$_{14}$. The novel hybrid allotropes were found stable mechanically (elastic constants and their combinations) and dynamically (phonons band structures). The thermal properties from the temperature dependence of the heat capacity C$_V$ were found to increasingly depart from diamond-like C$_8$ to higher values. From the electronic band structures, the inserted carbons were found to add up bands rigidly to diamond-like C$_8$ while being characterized by metallic-like behavior for ene-C$_{10}$ and ene-yne-C$_{14}$.*

**Introduction and context**

Original carbon allotropes, especially diamond-like ones, constitute a challenging and dynamic research field, not only for applications such as in power electronics, RT quantum computing, bio-sensors, bio-interfaces, color centers, etc... (cf. [1] review), but also at the fundamental research level thanks to modern structure research software such as USPEX [2] based on evolutionary crystallography and CALYPSO [3] based on particle swarm optimization.

Predicted carbon allotropes are assembled in the SACADA database [4] and organized with categories of topology determined by the TopCryst program [5], ex. **dia** for diamond, **lon** for 'Lonsdaleite' (a rare hexagonal form of diamond), labeled as 3C and 2H polytypes, respectively. Structural polytypes also characterize silicon carbide SiC, with the most well-known **cfc** 4H Moissanite variety. **cfc** topology is also adopted by hexagonal $C_8$ called "carbon 4C" (SACADA Database Code: 111, and ref. [6]).

In 3C, 2H, 4H, and other diamond polytypes, carbon hybridization is purely tetrahedral $C(sp^3)$ with C-C simple bonds characterizing ultra-hard large band gap insulating electronic systems. For application in electronics, diamond electronic structure, and subsequent physical properties can be modified by introducing carbon with different hybridizations as $C(sp^2)$ with C=C double bonds, found for instance in Nanodiamond where $sp^2/sp^3$ mixed carbon hybridization plays an important role in the design of advanced electronic materials [7]. We have also reported on the effects of increasingly smaller ratios $C(sp^2) / C(sp^3)$ on the electronic and mechanical structures of extended carbon networks based on diamond polytypes found to perform mechanically close to diamond [8]. Semi-metallic $C_{12}$ with mixed carbon hybridizations: $sp^3$ -$sp^1$ was identified as super-hard and called *yne*-diamond [9] where the "*yne*" prefix refers to C≡C triple bonds as in acetylene (also listed in SACADA Database Code: 112).

In this work, we develop the known hexagonal carbon allotropes having **cfc** topology, namely $C_8$, and $C_{12}$, by introducing original $C_{10}$ and $C_{14}$ characterized respectively by $sp^2/sp^3$ and $sp^1/sp^2/sp^3$ mixed carbon hybridizations. Following the above labeling of *yne*-diamond [9], we call them "*ene*"-$C_{10}$ with C=C double bonds, and "*ene-yne*"-$C_{14}$ characterized by both double C=C and triple C≡C bonds. However, in regard to the *yne*-diamond label, it will be shown that the hybrid allotropes possess lower hardness.

The investigations based on crystal chemistry 'engineering' are quantitatively supported by computations within the quantum mechanics framework of the Density Functional Theory (DFT) [10,11], reporting original results regarding the structure's stabilities, the mechanical and dynamic properties, as well as the electronic band structures.

1- **Computational framework**

The devised structures were all submitted to geometry relaxations of the atomic positions and the lattice constants down to the ground state characterized by minimal energy. The iterative computations were performed using DFT-based plane-wave Vienna Ab initio Simulation Package (VASP) [12,13]. For the atomic potentials, the projector augmented wave (PAW) method was used [13,14]. The exchange X and correlation effects (XC) were treated within a generalized gradient approximation scheme (GGA) [15]. Preliminary calculations using hybrid functional HSE06 [16] did not lead to better results versus GGA. The relaxation of the atoms onto ground state geometry was done by applying a conjugate-gradient algorithm [17]. Blöchl tetrahedron method [18] with corrections according to Methfessel and Paxton scheme [19] were applied for geometry optimization and energy calculations, respectively. A special $k$-point sampling [20] was applied for approximating the reciprocal space Brillouin-zone (BZ) integrals. For better reliability, the optimization of the structural parameters was carried out along with successive self-consistent cycles with increasing k-mesh until the forces on atoms were less than 0.02 eV/Å and the stress components below 0.003 eV/Å$^3$.

The mechanical stabilities and hardness were inferred from the calculations of the elastic constants [21]. The phonon dispersion band structures were calculated to verify the dynamic stability of the carbon allotropes. The phonon modes were computed considering the harmonic approximation through finite displacements of the atoms around their equilibrium positions to obtain the forces from the summation over the different configurations. The phonon dispersion curves along the direction of the Brillouin zone were subsequently obtained using "Phonopy" interface code based on Python language [22]. The crystal information files (CIF) and the structure sketches including the tetrahedral representations were produced with VESTA graphic software [23]. The electronic band structures and density of states were obtained with the full-potential augmented spherical wave ASW method based on DFT using the same GGA scheme as above [24].

## 2- Crystal chemistry and charge density illustration

**cfc**-$C_8$ structure shown in Fig. 1a, differs somehow from **lon** topology 4H $C_8$ (Fig. 1b) with a different stacking of the *C4* tetrahedra exhibited by the respective tetrahedral representations. Both allotropes crystallize in the highest symmetry hexagonal space group *P6₃/mmc* N° 194. From Table 1, the presently calculated crystal parameters as those from the SACADA #111 data show good agreements for the lattice constants and the atomic positions. The total energy and the atom-averaged cohesive energy are given in the last line. E(coh.)/at. eV= -2.49 eV is the same as for diamond and Lonsdaleite, letting us expect similar physical and electronic properties as shown for the 4H, 6H, and 8H polytypes (cf. [8] and cited works).

The topology and the space group are preserved upon inserting two additional carbon atoms at cell corners and hexagonal axis middle (0, 0, ½), i.e., at (2*a*) positions and then relaxing the structure through successive runs with increasing BZ integration in k-space. The resulting $C_{10}$ (called *ene*-$C_{10}$) is shown in Fig. 2a exhibiting aligned C-C-C along c-axis leading to larger volume per atom (5.97 Å$^3$) versus 5.65 Å$^3$ in $C_8$ and subsequent lower compactness as observed from the lowering of the density from 3.53 g/cm$^3$ down to 3.34 g/cm$^3$. The shortest C-C distance is now 1.46 Å along linear C—C—C. The cohesive energy is lowered with respect to **cfc-**$C_8$ due to the perturbation of the diamond-like structure with $sp^2$-like additional carbon.

Regarding the introduction of $sp^1$-carbon within *yne*-$C_{12}$ our calculations (Table 1) show good agreement with SACADA Database #112 and ref. [9]. The structure reproduced in Fig. 2b shows the effect of the triple C≡C bonds on the separation of the tetrahedra leading to an open structure. The density is lowered further with respect to $C_{10}$ and $C_8$: ρ(yne-$C_{12}$) = 2.99 g/cm$^3$.

Devising ene-yne-$C_{14}$ was subsequently operated to propose an original mixed tri-hybridized allotrope, with $sp^3/sp^2/sp^1$. The structure is shown in Fig. 2c and the crystal data (Table 1) feature distances of both double C=C bonds and triple C≡C bonds besides single bonds C-C, the latter forming the tetrahedra that are now more separated than in $C_{10}$ and $C_{12}$. The density is found close to 3 g/cm$^3$, meaning that ρ reached a minimum with yne-$C_{12}$ and decreases no more. The shortest distances are now 1.24 Å for C≡C and 1.39 Å for C=C. To the best of our knowledge, it's the first structure presenting the three kinds of hybridizations and we published and recorded it after curation at Cambridge Crystallographic Data Centre (CCDC) with ref. [25].

All three hybrid allotropes were found to belong to **cfc** topology.

Illustration of the changes from C-C to C=C and then to C≡C can be obtained from the projections of the charge densities on the crystal sites represented in Figure 3 with yellow volumes. For $C_8$, Fig. 3a features tetrahedral-shaped yellow volumes on all atoms, representing purely $sp^3$ hybridization. Turning to *ene*-$C_{10}$ (Fig. 3b), besides tetrahedral volumes there are now linear shapes along the c-hexagonal vertical direction of linear C—C—C signaling $sp^2$-like hybridizations like in $C_3H_4$ propadiene molecule (H2>C=C=C<H2; the ">" and "<" sign symbolizing bonds C with two H). In *yne*-$C_{12}$ (Fig. 3c), the charge density along C—C—C—C shows large yellow volumes around the central C-C signaling C≡C $sp^1$-like hybridizations. Lastly, in *ene-yne*-$C_{14}$ (Fig. 3d), $sp^3$, $sp^2$, and $sp^1$-like volumes can be observed. Then, such projections provide accurate illustrations of C…C different types of bindings.

3- **Mechanical properties from the elastic constants**

The analysis of the mechanical behavior was subsequently carried out using the elastic properties by performing finite distortions of the lattice. The chemical system is then fully described by the bulk (*B*) and the shear (*G*) moduli obtained by an averaging protocol of the elastic constants. Herein, we used Voigt's method (cf. [21] for original and modern works), based on a uniform strain. The calculated sets of elastic constants $C_{ij}$ (i and j correspond to directions) are given in Table 2. All $C_{ij}$ values are positive. The elastic constants of $C_8$ have the largest magnitudes, close to diamond [26]. Structurally related allotropes $C_{10}$, $C_{12}$, and $C_{14}$ show a trend of decreasing magnitudes. The four allotropes can be differentiated from the bulk $B_V$ and shear $G_V$ moduli, obtained from the equations corresponding to the hexagonal system [27]:

$B_V = 1/9 \{2(C_{11} + C_{12}) + 4C_{13} + C_{33}\}$

$G_V = 1/30 \{C_{11} + C_{12} + 2C_{33} − 4C_{13} + 12C_{44} + 6(C_{11} − C_{12})\}$

The last columns of Table 2 provide the obtained $B_V$ and $G_V$. $C_8$ has the largest values, close to the accepted values for diamond $B_V$ =445 GPa and $G_V$ = 550 GPa [26]. The other allotropes also have large bulk and shear moduli values, close to diamond, especially *ene*-$C_{10}$. Lastly,

*ene-yne*-C$_{14}$ shows the smallest magnitudes. Such trends are expected to reflect the hardness properties.

The last column of Table 2 provides the Vickers hardness ($H_V$) obtained with Chen et al. semi-empirical model [28]. Expectedly, C$_8$ has the largest magnitude with 90 GPa, close to diamond's [26] and the three hybrid allotropes show decreasing values along the series. The value of hardness of *yne*-C$_{12}$ $H_V$= 59 GPa [9] evaluated using the Gao et al. model [29] shows a close magnitude to the presently calculated value of $H_V$= 57 GPa.

Then the decrease of hardness along the series follows closely the trend of decreasing density concomitant with the openness of the respective structures arising from the separating of *C4* tetrahedra caused by the introduction of sp$^2$ and sp$^1$ -hybridized carbon atoms.

### 4- Dynamic properties from the phonons

A relevant criterion of phase stability is obtained from the phonon's properties. Phonons described as quanta of vibrations have their energy quantized thanks to the Planck constant 'h' used in its reduced form ℏ (ℏ = h/2π). The phonons energy: E = ℏω (frequency: ω) is then obtained and the corresponding band structures are plotted along the Brillouin zone in the reciprocal space.

All four carbon allotropes were submitted to the investigation of the phonons to determine their respective dynamic properties. Fig. 4 shows the phonon bands. Along the horizontal direction, the bands develop along the main lines of the hexagonal Brillouin zone (reciprocal k- space). The vertical direction shows the frequencies ω given in units of terahertz (THz).

There are 3N-3 optical modes found at higher energy than three acoustic modes that start from zero energy (ω = 0) at the Γ point, center of the Brillouin Zone (BZ), up to a few Terahertz. They correspond to the lattice rigid translation modes with two transverse and one longitudinal. The remaining bands correspond to the optic modes. In the four panels, there are no negative frequencies, and the corresponding allotropes are dynamically stable. In **cfc** C$_8$, the highest band culminates at ω ~ 40 THz, a magnitude observed for diamond by Raman spectroscopy [30]. Higher energy bands characterize the three hybrid allotropes: in C$_{10}$ at ω$_{max}$ ~ 42 THz assigned to antisymmetric C-C-C stretching in allene (propadiene) molecule (cf. [8]. The phonons band structure of C$_{12}$ exhibits flat bands at 70 THz that can be assigned to C≡C stretching; such higher frequency arises from the much shorter inter-carbon distance than in C=C. Lastly, for C$_{14}$ with sp$^1$ and sp$^2$ carbons, intermediate values are observed. The phonon band structures are found to provide a quantitative explanation of the chemical behaviors.

### 5- Thermal properties

The thermal properties as entropy and heat capacity $C_v$ were calculated using the statistical thermodynamic expressions from the phonon frequencies on a high precision sampling mesh in the BZ (cf. the textbook by Dove on 'Lattice Dynamics' [31]). Focusing on the diamond-like $C_8$, Fig. 5a shows the temperature change of entropy and heat capacity at constant volume. Along the vertical axis, the unit is Joule.atom$^{-1}$.K$^{-1}$, i.e., as per one carbon atom. Entropy S increases regularly with T. $C_v$ experimental data on diamond's heat capacity up to high temperatures were early obtained back in 1962 [32]; the experimental points plotted as red-filled circles on the calculated green curve show an excellent agreement with $C_8$ calculated curve. In the second step, it became interesting to compare the heat capacity behavior of the other allotropes with $C_8$ and diamond. Fig. 5b shows a continuous departure from $C_8$ (and diamond) to higher CV magnitudes along $C_{10}$, $C_{12}$, and then $C_{14}$. Such behavior is concomitant with the change of local carbon hybridization inserted into the diamond-like **cfc** $C_8$ lattice; i.e., from $C(sp^2)$ in $C_{10}$ to $C(sp^1)$ in $C_{12}$, and lastly original mix of $C(sp^2/sp^1)$ in $C_{14}$. The behavior of $C(sp^2)$ in $C_{10}$ resembles closely the results obtained in the series $C_{10}$, $C_{14}$, and $C_{18}$ with decreasing amounts of $C(sp^2)$ doping [8].

### 6- Electronic band structures and density of states

Using the crystal parameters in Table 1, the electronic band structures were obtained using the all-electrons DFT-based augmented spherical method (ASW) [24] and shown in Figure 6. The bands develop along the main directions of the primitive tetragonal Brillouin zones. In so far that $C_8$ and $C_{12}$ band structures are characterized by band gaps between the valence band (VB) and the empty conduction band (CB), the energy reference along the vertical energy axis is with respect to the top of the VB. Note that whereas $C_8$ is insulating with a large band gap like diamond, $C_{12}$ is semi-conducting. $C_{10}$ and $C_{14}$ are metallic with bands crossing the Fermi level $E_F$. The passage from metal $C_{10}$ to semi-conductor $C_{12}$ can be explained by the larger localization of the electron density from $sp^2$ to $sp^1$ as shown in Fig. 3. Lastly the VB band structure shape characterizing $C_8$ is reproduced in the three other hybrid allotropes and well-separated from the bands corresponding to the hybrid carbons, most obviously marked in $C_{14}$.

Such observation lets confirm that the $sp^2$ and $sp^1$ carbons are interstitials adding up to the diamond-like lattice, little perturbing it.

## Conclusions and prospectives

**cfc** hybrid carbon allotropes made of inserting $C(sp^2)$ and $C(sp^1)$ behaving carbon atoms into the diamond-like lattice of $C_8$ were identified and subsequently labeled *ene*-$C_{10}$ containing $C(sp^2)$ and *ene-yne*-$C_{14}$ containing $C(sp^2$ and $sp^1)$. The resulting changes brought by the double C=C and triple C≡C were illustrated from the projected charge densities. The density and the cohesive energy were found to decrease along the series due to the larger openness of the structures due to inserted $sp^2/sp^1$ carbons with decreasing hardness. The novel hybrid allotropes were found mechanically and dynamically stable. Changes in the electronic band structures due to the inserted carbons were found to add up rigidly to the diamond-like bands of $C_8$ while bringing semi-conducting to metallic-like behaviors. Modified diamonds such as modeled herein, exist in nanodiamonds. Their investigations induce growing interest in materials sciences communities. Applications are envisaged as electrochemical ones [7].

Besides the presented results confirming the cohesiveness, the stability (both mechanically and dynamically), as well as a full description of the electronic properties confirming findings on the novel carbon systems all concluding to the potential existence of such allotropes, this work can be further completed with molecular dynamics simulations to check the kinetic stabilities. Such heavy computations require improved computational means (software and hardware) not available now.

**TABLES**

Table 1 Crystal structure parameters of carbon allotropes with **cfc** topology. Values between brackets are those calculated in present work for $C_8$ and $C_{12}$

| $P6_3/mmc$ N°194 | **cfc** $C_8$ SACADA Database #111 | *ene* $C_{10}$ | *yne* $C_{12}$ SACADA Database #112 | *yne-ene* $C_{14}$ |
|---|---|---|---|---|
| $a$, Å | 2.513(2.512) | 2.487 | 2.623(2.634) | 2.609 |
| $c$, Å | 8.266(8.283) | 11.152 | 13.338(13.320) | 16.106 |
| $V_{cell}$, Å$^3$ | 45.209(45.246) | 59.739 | 79.461(80.019) | 94.94 |
| $<V_{atom}>$ Å$^3$ | 5.65 (5.66) | 5.97 | 6.62 (6.67) | 6.78 |
| Density $\rho$ g/cm$^3$ | 3.53 | 3.34 | 2.99 | 3.02 |
| Shortest d-d- Å | 1.54 | 1.46, 1.54 | 1.20, 1.44, 1.53 | 1.24, 1.39, 1.41 |
| Atomic position | C1 (4e) 0,0,0.0929(0.093) C2(4f) 2/3 1/3 0.1553 (0.156) | C1(2a) 0,0,0 C2 (4e) 0,0,0.369 C3(4f) 2/3 1/3 0.181 | C1(4e) 0,0,0.557(0.557) C2(4f) 2/3 1/3 0.597(0.597) C3(4f) 2/3 1/3 0.7049(0.705) | C1(2a) 0,0,0 C2 (4e) 0,0,0.588 C3(4f) 2/3 1/3 0.624 C4(4f) 2/3 1/3 0.712 |
| $E_{(Total)}$ eV | -72.68 | -85.0 | 100.43 | 113.23 |
| E(coh.)/at. eV | -2.49 | -1.90 | -1.76 | -1.49 |

Table 2. Elastic constants $C_{ij}$ and Voigt values of bulk ($B_V$) and shear ($G_V$) moduli (all values are in GPa). Diamond-like $C_8$ and hybrid $C_{10}$, $C_{12}$, and $C_{14}$ *ene, yne* carbon allotropes with **cfc** topology. The hardness value between brackets is from ref. [**9**]

|  | $C_{11}$ | $C_{12}$ | $C_{13}$ | $C_{33}$ | $C_{44}$ | $B_V$ | $G_V$ | $H_V$ |
|---|---|---|---|---|---|---|---|---|
| $C_8$ | 1188 | 105 | 37 | 1267 | 541 | 445 | 513 | 90 |
| *ene* $C_{10}$ | 900 | 121 | 35 | 1498 | 389 | 409 | 440 | 74 |
| *yne* $C_{12}$ | 617 | 45 | 66 | 1353 | 286 | 327 | 332 | 57(59) |
| *ene-yne* $C_{14}$ | 499 | 124 | 57 | 1666 | 187 | 349 | 276 | 38 |

**FIGURES**

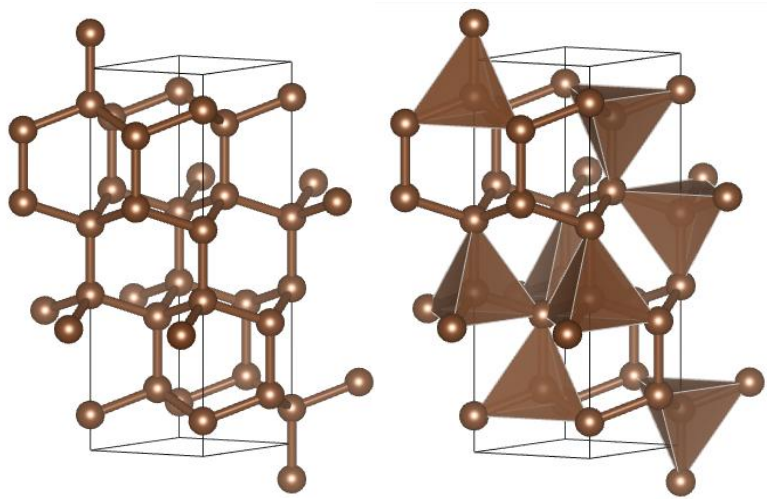

a) **cfc** $C_8$

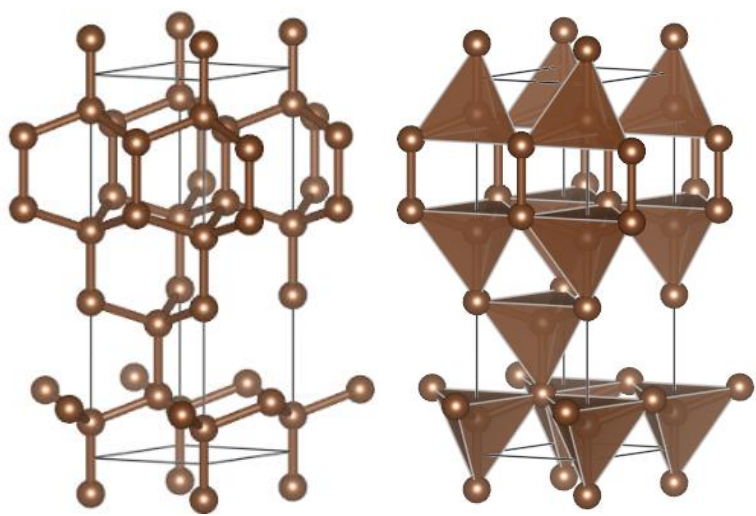

b) **lon** $C_8$ (4H)

Figure 1. Crystal structures of hexagonal $C_8$: a) **cfc** (4C carbon -SACADA #111) b) Lonsdaleite, with ball-and-stick and polyhedral representations highlighting the stacking of *C4* tetrahedra.

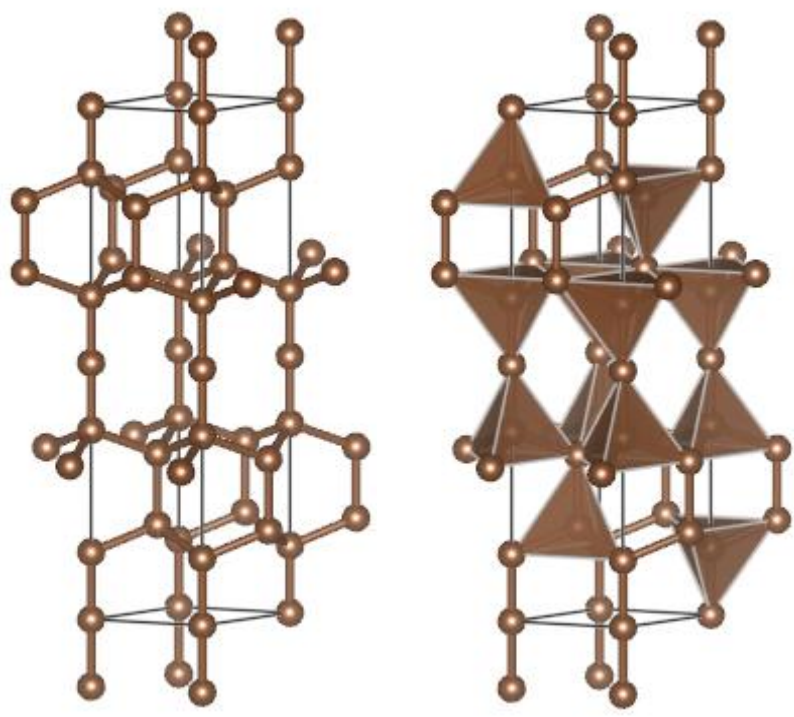

a) *ene*-C$_{10}$

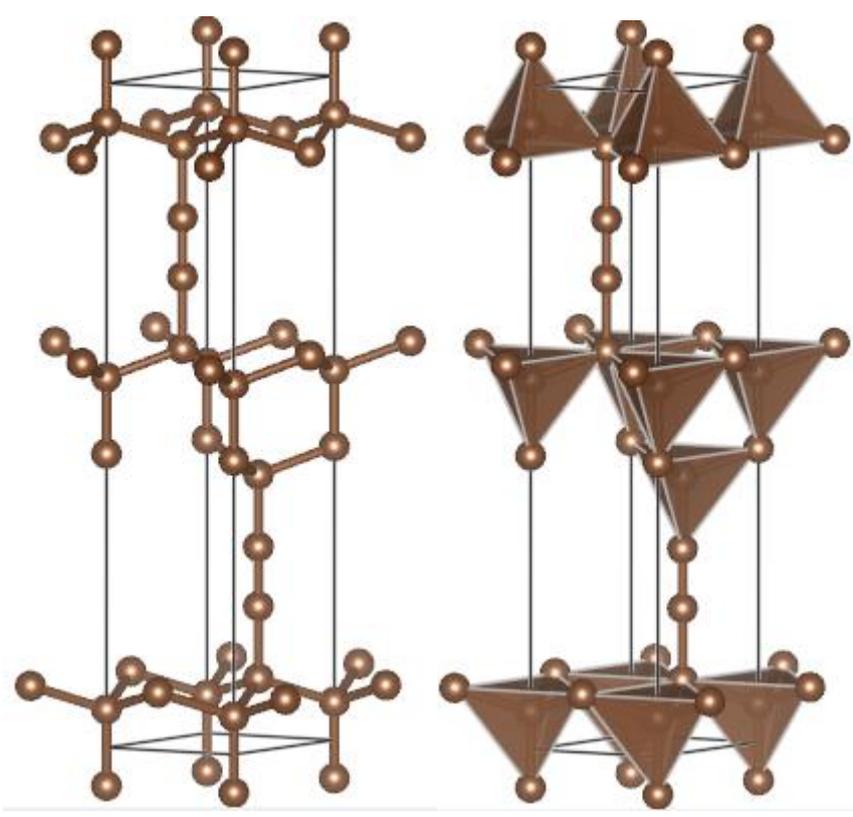

b) *yne*-C$_{12}$[9]

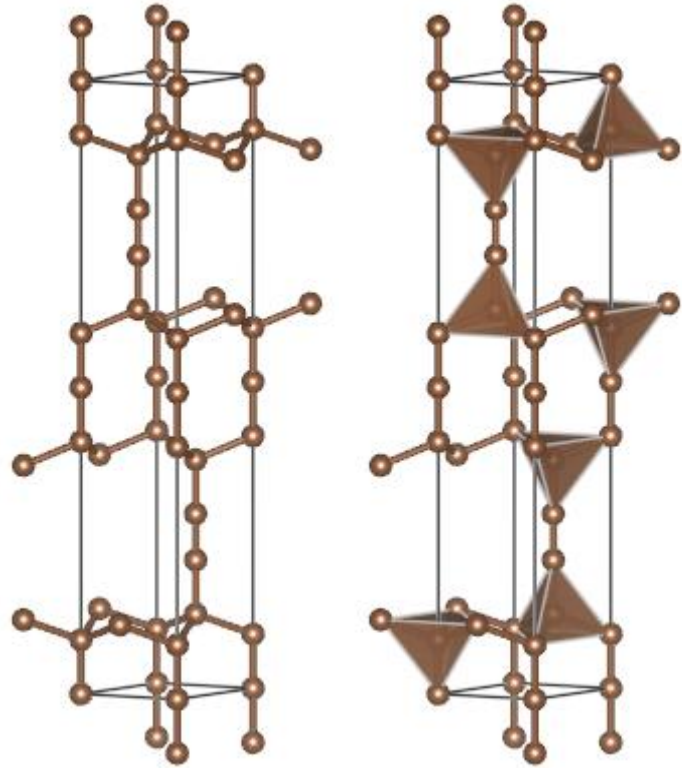

c) *ene-yne*-C$_{14}$

Figure 2. Crystal structures of hybrid carbon allotropes with **cfc** topology (ball-and-stick and polyhedral representations highlighting the *C4* tetrahedral stacking). a) *ene*-C$_{10}$, b) *yne*-C$_{12}$[9], c) *ene-yne*-C$_{14}$

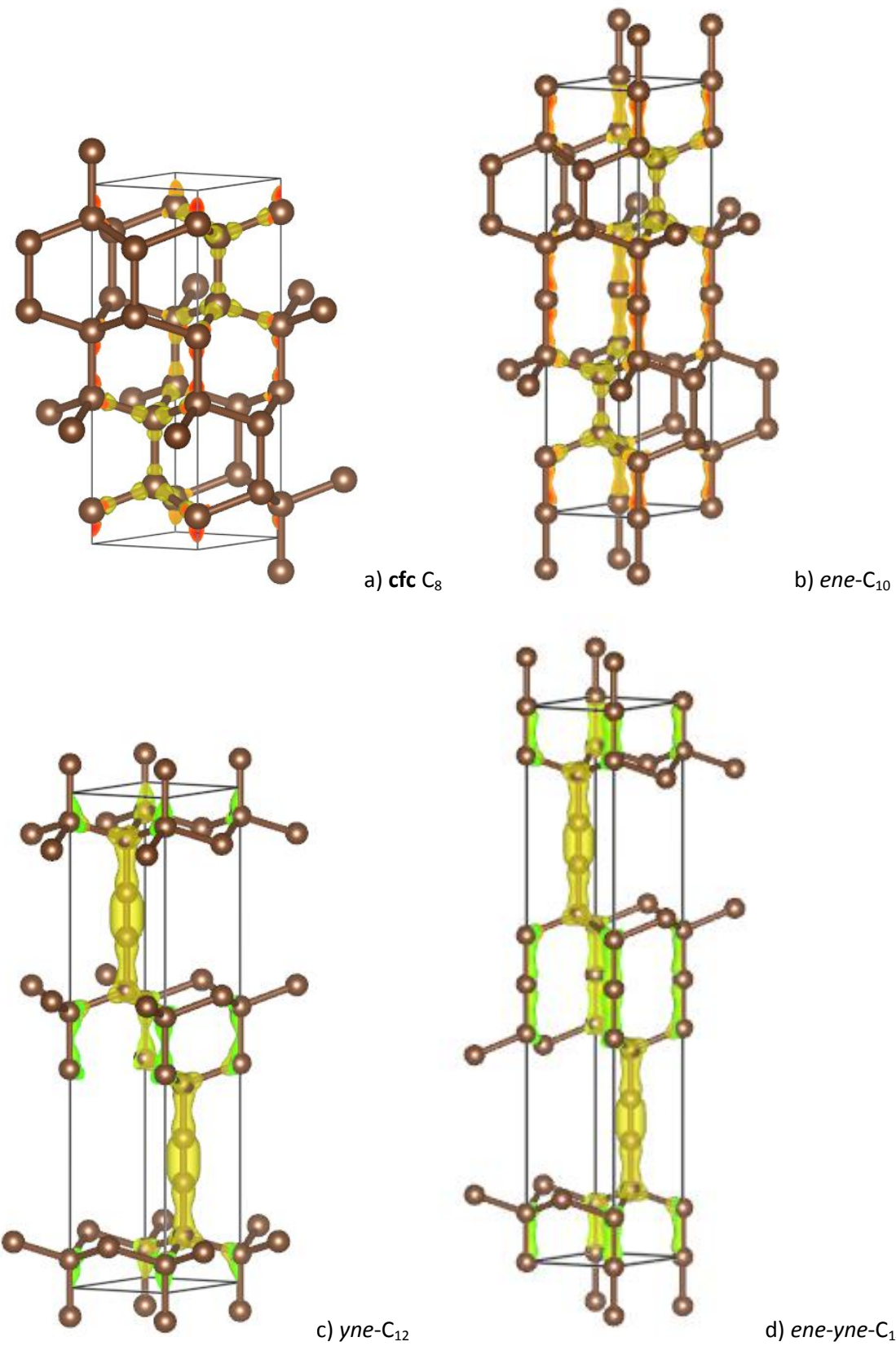

Figure 3. Charge density projections (yellow volumes) a) $C_8$, b) *ene*-$C_{10}$, c) *yne*-$C_{12}$[9], d) *ene-yne*-$C_{14}$

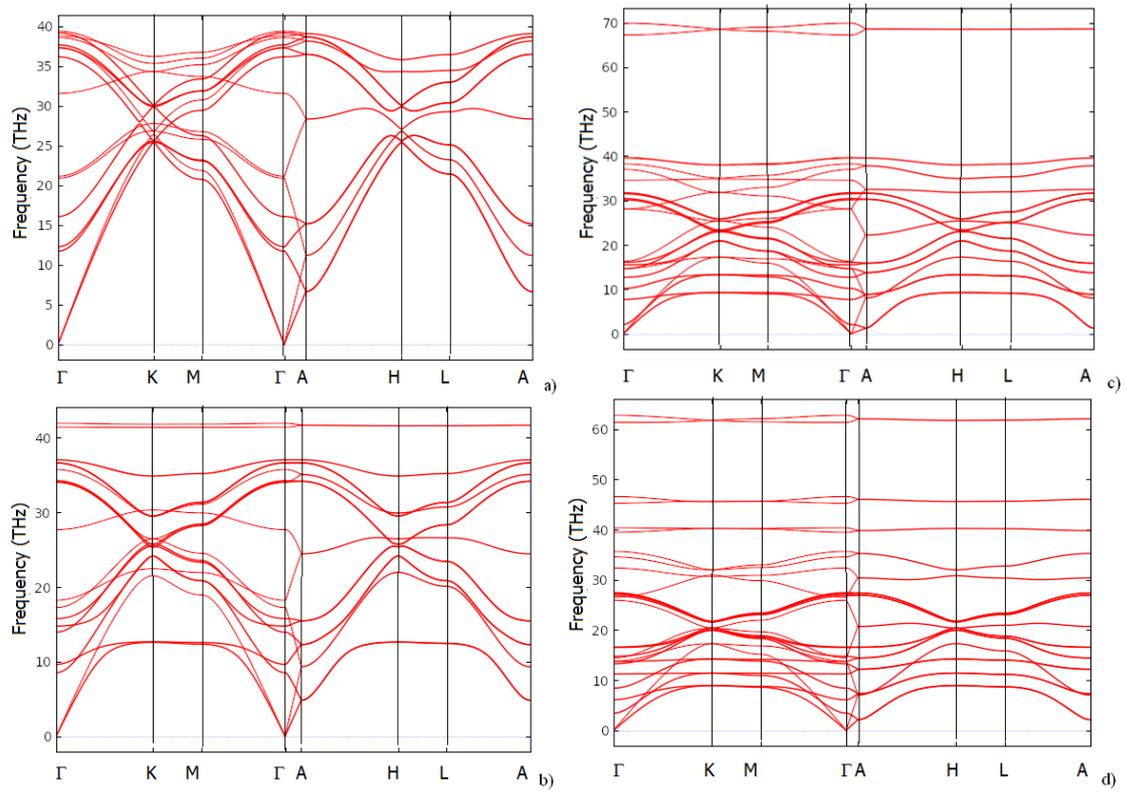

Figure 4. Phonon band structures: a) $C_8$, b) *ene*-$C_{10}$, c) *yne*-$C_{12}$, d) *ene-yne*-$C_{14}$

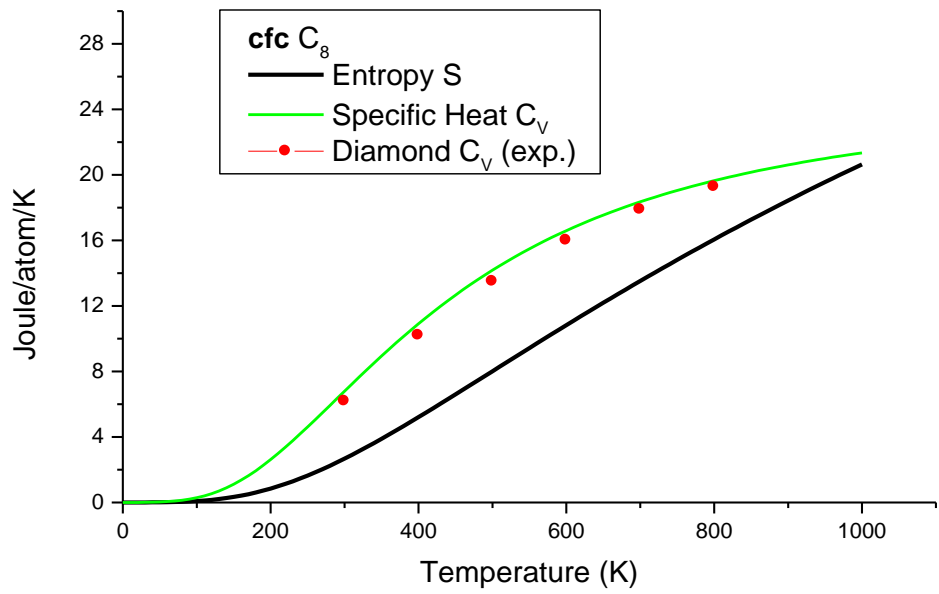

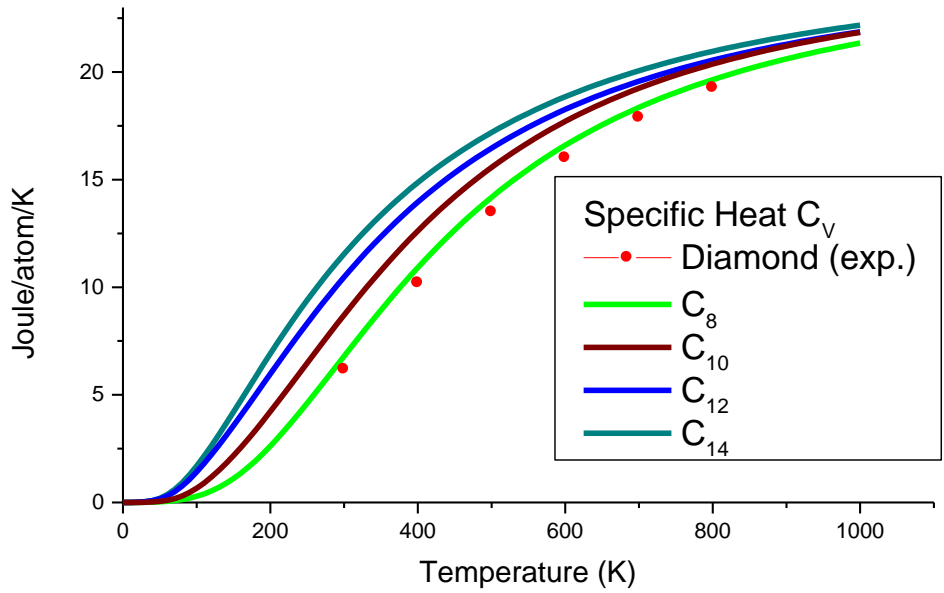

Figure 5. Temperature evolution curves of the entropy and specific heat in **cfc** $C_8$, with experimental points of diamond $C_V$. b) $C_V = f(T)$ in $C_8$, $C_{10}$, $C_{12}$, and $C_{14}$

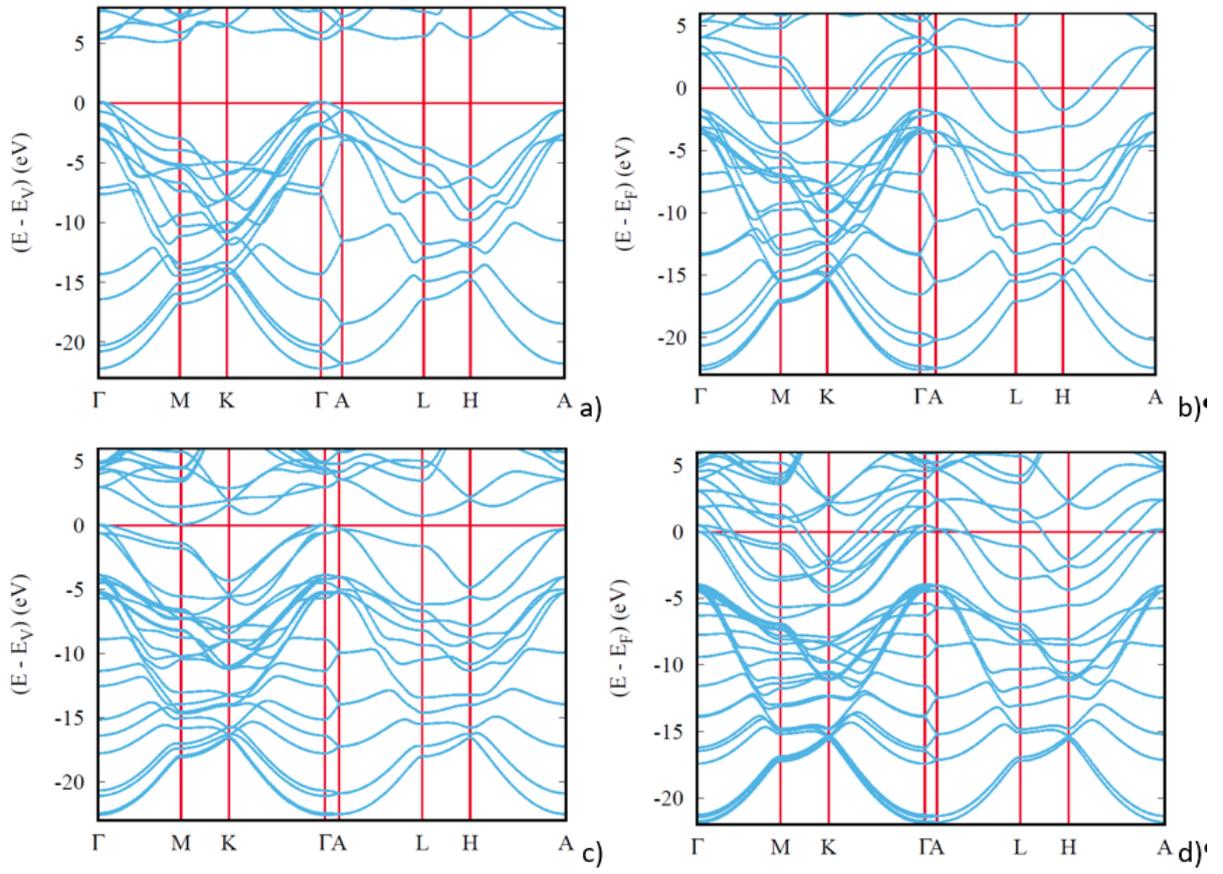

Figure 6. Electronic band structures a) $C_8$, b) *ene*-$C_{10}$, c) *yne*-$C_{12}$, d) *ene-yne*-$C_{14}$